\documentclass[12pt]{iopart}
\usepackage{iopams}

\def\be{\begin{equation}}
\def\ee{\end{equation}}

\begin{document}
\title{Noether symmetry in the higher order gravity theory}
\author{A.K. Sanyal $^{a, \dag}$\,
 B.Modak\,$^{c}$ C. Rubano\,$^{d,\ddag}$\, E. Piedipalumbo \,$^{d,\ddag}$
 \footnote[3]{electronic-mail:\\
 author for correspondence: ester@na.infn.it\\ aks@juphys.ernet.in,\\
bmodak@klyuniv.ernet.in\\rubano@na.infn.it }}

\address{$^a$  Department of Physics, Jangipur College,
Murshidabad, India - 742213}
\address{$^\dag$ Cosmology Research Centre, Dept. of Physics, Jadavpur University, Kolkata - 700032, India}
\address{$^c$ Department of Physics, University of Kalyani, Kalyani, 741235, India}
\address{$^d$ Department of Physics, University of Napoli, Naples,
Italy}
 \address{$^\ddag$INFN sezione di Napoli, Complesso di MSA, via Cintia 80126,  Napoli, Italia }

\begin{abstract}
Noether symmetry for higher order gravity theory has been explored, with
the introduction of an auxiliary variable which gives the only correct
quantum desccription of the theory, as shown in a series of earlier
papers. The application
of Noether theorem in higher order theory of gravity turned out to be a
powerful tool to
find the solution of the field equations. A few such physically
reasonable solutions like power law inflation are presented.

\end{abstract}
\maketitle
\section{\bf{Introduction}}
The higher order gravity theory has important contribution in the early
universe. The importance of fourth order gravity in the gravitational action
was explored by several authors. Starobinsky \cite{s:plb} first presented a solution of the
inflationary scenario without invoking phase transition in the early universe
considering only geometric term in the field equations. In this direction,
Hawking and Luttrel \cite{hl:prl} shown the curvature square term in the action
mimicks as
a massive scalar field. Further, Starobinsky and Schmidt \cite{ss:cqg} have shown that the
inflationary phase is an attractor in the neighbourhood of the solution of
the fourth order gravity theory. Low energy effective action corresponding to Brane
world cosmology also contains higher curvature invariant terms.
\par
   To elucidate the effect of the fourth order gravity theory in the early
universe one needs more (exact) solutions of the field equations. The
solutions
of the classical fourth order gravity theory are very few due to non-linearity
of the field equations. The field equations obtained from the action can be
simplified by introducing an auxiliary variable following the prescription
of
Boulware et al \cite{b:a}.
\par
In some recent papers \cite{a:p}, \cite{a:cqg}, \cite{a:pl} the
minisuperspace
quantization of fourth order gravity
introducing auxiliary variable have been presented, whose canonical
quantization
yields Schrodinger like equation with a meaningful definition of quantum
mechanical probability. Further, the extremization of the effective potential
leads to the vacuum Einstein equation. The choice of auxiliary variable is the
turning point in simplifying the field equations and it yields a transparent
and simple quantum mechanical equation. With such a successful choice of
auxiliary variable it is extremly necessary to study the solution of the
classical field equations. The field equations with the auxiliary variable
are quite complicated and require special attention for solution. We introduce
Noether symmetry to the fourth order gravity theory in the FRW background
as
simplifying assumption instead of any adhoc assumption, or any equation of
state
for solution of the classical field equations.
\par
we consider the Einstein-Hilbert action with a curvature squared term and
a non-minimally coupled scalar field. Applying the Nother symmetry in the
above action following the Noether symmetry approach of de-Rities et al
\cite{de:prd}
it is possible to extract a class of solutions. It is interesting to note
that
the introduction of Noether symmetry in the action of higher order gravity
theory not only gives Noether symmetry but it also leads explicit time
dependence of the scale factor (as well as the scalar field)  as a consequence
of Noether symmetry.

We now move on to extract Noether
symmetry of the theory. It was earlier attempted by Capozziello
\cite{c:l},
where he introduced the higher order term in the action as a constraint,
through Lagrange multiplier. We shall try to compare the results also.

\section{\bf{Classical field equations and the equations governing Noether
symmetry}}
We consider the following action.
\be
S=\int d^4 x\sqrt{-g}[\frac{1}{16\pi
G}(f(\phi)R+\frac{\beta}{6}R^2)-\frac{1}{2\pi^2}(\frac{1}{2}
\phi,_{\mu}\phi'^{\mu}+V(\phi))]
\ee
 In the Robertson-Walker metric
\be
ds^2=e^{2\alpha}[-d\eta^2+d\chi^2+F(\chi)(d\theta^2+sin^2\theta d\phi^2)]
\ee

The action takes the form,
\begin{eqnarray}
&&S=\int[\frac{3\pi}{2G}\{f\alpha''
e^{2\alpha}+f(\alpha'^2+k)e^{2\alpha}+\beta\alpha''^2+\beta(\alpha'^2+k)^2+2\beta\alpha''(\alpha'^2+k)\}\nonumber\\
&&+\frac{1}{2}\phi'^2 e^{2\alpha}- V(\phi)e^{4\alpha}]d\eta.
\end{eqnarray}
In the above, dash(') denotes derivative with respect to $\eta$
and $k=0,$. Removing total derivative terms and choosing
$\frac{3\pi}{2G} = M$, the action can be expressed as,
\begin{eqnarray}
&&S=\int[M\{f(k-\alpha'^2)e^{2\alpha}-f,_{\phi}\alpha'\phi'e^{2\alpha}+\beta(\alpha'^2+k)^2
+\beta\alpha''^2\}\nonumber\\&&+\frac{1}{2}\phi'^2
e^{2\alpha}-V(\phi)e^{4\alpha}]d\eta+\Sigma_{1}.
\end{eqnarray}
Where
$\Sigma_{1} = M[f\alpha'
e^{2\alpha}+2\beta(\frac{\alpha'^3}{3}+k\alpha')]$ is the surface
term. Now we introduce the auxiliary variable $Q$ as, \be
MQ=\frac{\partial S}{\partial\alpha''}=2M\beta \alpha'',~~ ie.,
Q=2\beta\alpha''. \ee Introducing the auxiliary variable in the
action and writing the action in the canonical form we obtain,
\begin{eqnarray}
&&S=\int[M\{f(k-\alpha'^2)e^{2\alpha}-f,_{\phi}\alpha'\phi'
e^{2\alpha}+\beta(\alpha'^2+k)^2+Q\alpha''-\frac{Q^2}{4\beta}\}\nonumber\\&&+\frac{1}{2}\phi'^2
e^{2\alpha}-Ve^{4\alpha}]d\eta+\Sigma_{1}. \end{eqnarray}
Finally
removing total derivative term appearing in the auxiliary variable
we obtain,
\begin{eqnarray}
&&S=\int[M\{f(k-\alpha'^2)e^{2\alpha}-f,_{\phi}\alpha'\phi'e^{2\alpha}+\beta(\alpha'^2+k)^2-Q'\alpha'
-\frac{Q^2}{4\beta}\}\nonumber\\&&+\frac{1}{2}\phi'^2
e^{2\alpha}-Ve^{4\alpha}]d\eta+\Sigma, \end{eqnarray} where,
$\Sigma = \Sigma_{1}+MQ\alpha'$. It is not difficult to see that
the action is canonical, since, Hessian determinant
$|\Sigma\frac{\partial^2{L}}{\partial{q_{i}}'\partial{q_{j}'}}| =
- M^2 e^{2\alpha}\ne{0} $. Thus field equations are
\begin{eqnarray}&& 4\beta(3\alpha'^2
+k)\alpha''-2f(\alpha''+\alpha'^2+k)e^{2\alpha}-(\phi''
f,_{\phi}+2\alpha'\phi' f,_{\phi}+\phi'^2
f,_{\phi\phi})e^{2\alpha}\nonumber\\&&-Q'' =
\frac{1}{M}(\phi'^2-4V(\phi)e^{2\alpha})e^{2\alpha}.
\end{eqnarray}
\begin{eqnarray} Q = 2\beta\alpha''  \end{eqnarray}
\begin{eqnarray}
f,_{\phi}(\alpha''+\alpha'^2+k) =
\frac{1}{M}(\phi''+2\alpha'\phi'+V,_{\phi}e^{2\alpha}).
\end{eqnarray} Finally the Hamilton constraint equation is,
\begin{eqnarray}
&&[f(\alpha'^2+k)+f,_{\phi}\alpha'\phi']e^{2\alpha}-\beta(\alpha'^2+k)(3\alpha'^2-k)+Q'\alpha'
-\frac{Q^2}{4\beta}\nonumber\\&&=\frac{1}{M}[\frac{1}{2}\phi'^2+V(\phi)e^{2\alpha}]e^{2\alpha}.
\end{eqnarray} In the above system the configuration space is three
dimensional and its coordinates are $(\alpha,Q,\phi)$; whose
tangent space is specified by the variables
$(\alpha,Q,\phi,\alpha',Q',\phi')$. Hence we assume the
infinitesimal generator of the Noether symmetry as \be {\bf{X}} =
A\frac{\partial}{\partial\alpha} + B \frac{\partial}{\partial Q} +
C\frac{\partial}{\partial\phi}+
A'\frac{\partial}{\partial\alpha'}+ B'\frac{\partial}{\partial
Q'}+C'\frac{\partial}{\partial\phi'}, \ee where $A, B, C$ are
function of $\alpha,Q,\phi$. The existence of Noether symmetry
implies the existence of the vector field ${\bf{X}}$ such that the
Lie derivative of the Lagrangian with respect to the vector field
vanishes i.e. \be \pounds_{\bf{X}}L= 0. \ee The conserved quantity
corresponding to the Noether symmetry is \be F = A\frac{\partial
L}{\partial\alpha'}+ B\frac{\partial L}{\partial Q'} +
C\frac{\partial L}{\partial\phi'}. \ee equation (13) gives

\[
A[2M\{f(k-\alpha'^2)-f,_{\phi}\alpha'\phi'\}e^{2\alpha}+
(\phi'^{2}-4Ve^{2\alpha})e^{2\alpha}]\]
\[+(\frac{\partial A}{\partial \alpha}\alpha'+\frac{\partial A}{\partial
Q}Q' +\frac{\partial A}{\partial
\phi}\phi')M[(-2f\alpha'-f,_{\phi}\phi')e^{2\alpha}
+4\beta\alpha'(\alpha'^2+k)-Q']\]
\[+B(-\frac{MQ}{2\beta})
+(\frac{\partial B}{\partial \alpha}\alpha'+\frac{\partial B}{\partial Q}Q'
+\frac{\partial B}{\partial
\phi}\phi')(-M\alpha')\]
\[+C[M(f,_{\phi}(k-\alpha'^2)-f,_{\phi\phi}\alpha'\phi')
e^{2\alpha}-V,_{\phi}e^{4\alpha}]\]
\be
+(\frac{\partial C}{\partial \alpha}\alpha'+\frac{\partial C}{\partial Q}Q'
+\frac{\partial C}{\partial
\phi}\phi')(-Mf,_{\phi}\alpha' e^{2\alpha}+\phi' e^{2\alpha})=0
\ee

To satisfy equation (15) one has to satisfy the following equations:
Collecting co-efficients of $\alpha'^4, Q'\alpha'^3,$ and $\phi'\alpha'^3$,
we get
\be
A=A_{0}
\ee
where, $A_{0}$ is a constant. Co-efficient of $Q'\phi'$ gives
\be
\frac{\partial C}{\partial Q}=0,
\ee
ie., $C$ is not a function of $Q$. Further the co-efficients of $\alpha' Q'$
gives,
\be
f,_{\phi}\frac{\partial C}{\partial Q}+\frac{\partial B}{\partial Q}
e^{-2\alpha}=0,
\ee
ie., $B$ also does not depend on $Q$. Indeed it should be, since $Q$ is an
auxiliary variable only.
Co-efficient of $\phi'^2$ gives
\be
\frac{\partial C}{\partial \phi}+A=0,
\ee
which implies
\be
C=-A_{0}\phi+g_{1}(\alpha).
\ee
Finally, co-efficients of $\alpha'^2, \phi'\alpha'$ and $Q'\phi'$ give,
\be
2Af+f,_{\phi}(C+\frac{\partial C}{\partial \alpha})+\frac{\partial B}
{\partial \alpha}e^{-2\alpha}=0.
\ee
\be
f,_{\phi}(2A+\frac{\partial C}{\partial \phi})+Cf,_{\phi\phi}-\frac{1}{M}
\frac{\partial C}{\partial
\alpha}+\frac{\partial B}{\partial \phi}e^{-2\alpha}=0.
\ee
and
\be
k(2Af+Cf,_{\phi})e^{2\alpha}-\frac{BQ}{2\beta}-\frac{1}{M}(CV,_{\phi}+4AV)
e^{4\alpha}=0.
\ee
The solution of $A,B,C$ satisfying all these equations (16)-(23) yields
Noether symmetry.
\section{\bf{Solutions}}
Let us consider solution of equations (16)-(23) to find $A,B,C$ and hence
$f(\phi),V(\phi)$ admitting Noether symmetry,further we shall consider
solution of the field equations.
Choosing $B$ in the form $B = B_{1}(\alpha)B_{2}(\phi)$, given by the
separation of
variables, equation (21) gives,
\be
A_{0}(2f-\phi f,_{\phi})+f,_{\phi}(g_{1}+\frac{d g_{1}}{d\alpha}
+B_{2}\frac{dB_{1}}{d\alpha}e^{-2\alpha})=0.
\ee
Differentiating above equation (24) with respect to $\phi$ we get,
\be
A_{0}f,_{\phi}-A_{0}\phi f,_{\phi\phi}+(g_{1}+\frac{d
g_{1}}{d\alpha})f,_{\phi\phi}+B_{2,\phi}\frac{dB_{1}}{d\alpha}e^{-2\alpha}=0.
\ee
Eliminating, $g_{1}+\frac{dg_{1}}{d\alpha}$ between equations (23) and
(24) we get,
\be
A_{0}\frac{[(2f-\phi f,_{\phi}),_{\phi}f,_{\phi}-(2f-\phi
f,_{\phi})f,_{\phi\phi}]}{(f,_{\phi}B_{2,\phi}-B_{2}f,_{\phi\phi})}=
\frac{dB_{1}}{d\alpha}e^{-2\alpha}=N .
\ee
Since, left hand side is a function of $\phi$ and the right hand side is that
of $\alpha$, therefore
both sides are equated to a constant $N$.
Hence,
\be
B_{1}=\frac{N}{2}e^{2\alpha}+b_{0},
\ee
where $b_{0}$ is a constant, and
\be
2f-\phi f,_{\phi}=N_{1}f,_{\phi}-\frac{N}{A_{0}}B_{2},
\ee
$N_{1}$ being yet another constant. In view of equation (28), equation (24)
is,
\be
g_{1}+\frac{dg_{1}}{d\alpha}+A_{0}N_{1}=0,
\ee
for $f,_{\phi} \ne {0}$. Hence $g_{1}$ can be solved to find $C$ as,
\be
C=\alpha_{0}e^{-\alpha}-A_{0}(\phi+N_{1}).
\ee
In view of which the equation (22) takes the following form,
\begin{eqnarray}
&&A_{0}[f,_{\phi}-(\phi+N_{1})f,_{\phi\phi}]+\alpha_{0}(f,_{\phi\phi+
\frac{1}{M}})e^{-\alpha}\nonumber\\&&
+(\frac{N}{2}e^{2\alpha}+b_{0})e^{-2\alpha}B_{2,\phi}=0.
\end{eqnarray} This equation (31) is satisfied, provided
$\alpha_{0}=0$ and $b_{0}$ or $B_{2,\phi}$ $=0$. Note that, for
$f,_{\phi\phi}+\frac{1}{M}=0$, $f<0$, which leads to negative
Newton's gravitational constant. Now, for the first choice, ie.,
$\alpha_{0}=b_{0}=0$, the above equation (31) reads, \be
A_{0}[f,_{\phi}-(\phi+N_{1})f,_{\phi\phi}]+\frac{N}{2}B_{2,\phi}=0.
\ee Comparing equation (32) with the relation between $f$ and
$B_{2}$, given by equation (28) being differentiated with respect
to $\phi$ implies that these two equations are consistent either
for $N = 0$ or for $B_{2} =$ a constant. The first choice leads to
inconsistency. So finally we are left with only one option, ie.,
$\alpha_{0} = 0 = B_{2,\phi}$, ie., $B_{2} = b_{2}$, a constant.
For this choice equation (31) is \be
f,_{\phi\phi}(\phi+N_{1})-f,_{\phi}=0. \ee Further, equation (28)
gives, \be (\phi+N_{1})f,_{\phi}=2f+\frac{Nb_{2}}{2A_{0}}. \ee
Equations (33) and (34) are thus consistent and yield the
following solution, \be
f=f_{0}(\phi+N_{1})^2-\frac{Nb_{2}}{2A_{0}}, \ee along with \be
A=A_{0};~~B=b_{2}(\frac{N}{2}e^{2\alpha}+b_{0});~~C=-A_{0}(\phi+N_{1}).
\ee In view of (35, 36) equation (23) reads, \be
kNb_{2}e^{2\alpha}+\frac{b_{2}}{2\beta}(\frac{N}{2}e^{2\alpha}+b_{0})Q=
\frac{A_{0}}{M}[(\phi+N_{1})V,_{\phi}-4V]e^{4\alpha}. \ee Now
depending on values of the integration constants we consider the
following different cases:
\par
\subsection{\bf{Case 1, $b_{0} = 0,b_{2}\neq 0,(\phi+N_{1})V,_{\phi}=4V$.}}
\par
Under this situation equation (37) yields, \be Q=-4k\beta, ie.,
\alpha''=-2k.~~V(\phi)=V_{0}(\phi+N_{1})^4 \ee The scale factor
$e^{\alpha}$ can be obtained easily from equation (38) and it can
be used in the Noether constant of motion (14) to find solution
for $\phi$. Equation (14) takes the form
\begin{eqnarray}
&&\frac{F}{A_{0}M}= 4\beta\alpha'(k+\alpha'^2)-
Q'-\left[2f\alpha'+f_{,\phi}\phi'
+\frac{b_{2}N}{2A_{0}}\alpha'\right.\nonumber\\
&&-\left.(\phi+N_{1})\alpha'f_{,\phi}
+\frac{\phi+N_{1}}{M}\phi'\right] e^{2\alpha}. \end{eqnarray} To
find simple solution we choose $k=0$, then \be e^{\alpha}= e^{g
\eta}, \ee where $g$ is a constant and integration of equation
(39) yields \be (\phi+N_{1})^2 =\phi_{0}^2 e^{-2g\eta}
+\frac{C_{1}+b_{2}MNg/2}{A_{0}(1+2Mf_{0})/2}, \ee where $C_{1}$ is
a constant and $\phi_{0}^2=\frac{F/2g -2A_{0}M\beta g^2}
{A_{0}(1+2Mf_{0})/2}$. It is to be noted that the solution for
$\alpha$ and $\phi$ presented here are obtained from Noether
symmetry conditions and this solutions (40) and (41) satisfy the
field equations (8)-(10) trivially under a simple restriction on
the integration constants $C_{1}=-b_{2}MNg/2$ and
$V_{0}=\frac{g^2(1+2Mf_{0})}{4\phi_{0}^2}$. This solution
represents a power law inflation as the scale factor in proper
time is $e^{\alpha} = a_{0}t$.
\subsection{\bf{Case 2. $b_{0}=0, b_{2}\neq 0, (\phi+N_{1})V_{,\phi}-4V=r=\rm constant$}}
Equation (37) now  takes the form \be Q= 2\beta \omega_{0}^2
e^{2\alpha} - 4 k \beta, \ee \be V =V_{0} (\phi + N_{1})^4 + r,
\ee where $\omega_{0}^2 = \frac{2 A_{0}r}{MN b_{2}}$. Now using
(9) in (42) we get \be \alpha'^{2}= \omega_{0}^{2} e^{2\alpha} - 4
k \alpha + q^{2} \ee whose integral gives \be e^{- \alpha} =
\frac{\omega_{0}}{q} \sinh(q\eta) \ee where $q$ is an integration
constant. The solution (45) can be used in the Noether constant of
motion to find $\phi$ and is given by \begin{eqnarray}&& (f_{0} +
\frac{1}{2M})(\phi+N_{1})^{2} = \frac{F
\omega_{0}^{2}}{2Mq^{2}A_{0}} (\eta- \frac{sinh(
2q\eta)}{2q})\nonumber\\&& - 2\beta \omega_{0}^{2}sinh( q\eta^{2})
-\frac{Nb_{2}}{2A_{0}}\ln |sinh(q\eta)| + C_{2}, \end{eqnarray}
where $C_{2}$ is a constant. It is important to note that the
solutions (45) and (46) are obtained from the Noether symmetry
only. To justify consistency of the solution above $e^{\alpha}$
and $\phi$ have to satisfy the field equations (8)-(10).
Analytically this proof is too complicated so we leave this case.
\par
Another simple solution of equation for $k=0$ and $q=0$ is
\be
e^{- \alpha} = \omega_{0} \eta
\ee
and as a consequence solution of $\phi$ from (14) is
\be
(f_{0}+\frac{1}{2M})(\phi + N_{1})^{2} = C_{2} -\frac{F \eta^{3}}{3A_{0}M}
-\frac{Nb_{2}}{2A_{0} \omega_{0}^{2}}\ln\eta.
\ee
the solutions (47) and (48) obtained from the Noether symmetry is not
consistent with the field equations.
\subsection{\bf{Case 3.  f=\rm constant}}

It is also possible to study a totally different case viz., $f =$
constant $= 1$ (say).
\par
Under this assumption, equation (21) gives
\be
B=-A_{0}e^{2\alpha}+B_{2}(\phi).
\ee
Equation (22) is,
\be
MB_{2,\phi}=\frac{dg_{1}}{d\alpha}e^{2\alpha}=N,
\ee
where $N$ is the separation constant. Equation (50) is solved to yield,
\be
B_{2}=\frac{N}{M}\phi+B_{0};~~ g_{1}=-\frac{N}{2}e^{-2\alpha}+C_{0}.
\ee
Equation (23) thus becomes,
\begin{eqnarray}
&&A_{0}e^{2\alpha}[2k+\frac{Q}{2\beta}+\frac{N}{2MA_{0}}V,_{\phi}]-
\frac{N}{2M\beta}Q\phi\nonumber\\&&-\frac{B_{0}}{2\beta}Q+\frac{1}{M}[(A_{0}\phi-C_{0})
V,_{\phi}-4A_{0}V]e^{4\alpha}=0. \end{eqnarray} Now, the choice $Q
= Q_{0}e^{2\alpha}$ leads to inconsistency. The other choice is
$B_{0}=0=N$. Under this choice equation (50) gives, \be Q=-4\beta
k;~~ V=V_{0}(A_{0}\phi-C_{0})^4. \ee together with \be A=A_{0};~~
B=-A_{0}e^{2\alpha};~~ C=C_{0}-A_{0}\phi. \ee The conserved
current is, \be \frac{F}{A_{0}M}=4\beta
\alpha'(\alpha'^2+k)-Q'-\alpha'
e^{2\alpha}+\frac{C_{0}}{A_{0}M}\phi'
e^{2\alpha}-\frac{1}{M}\phi\phi' e^{2\alpha} \ee Now from (53) \be
\alpha = -k\eta^{2}+g\eta +h, \ee where $g,h$ are integration
constant. This solution (56) can be used in equation (55) to find
the scalar field and is given by \be
\phi^{2}=\frac{F}{gA_{0}}e^{-2g\eta}, \ee where we have assumed
$k=0$,$h=0$. Further, one has to check the consistency of
solutions (56) and (57) with the field equations. They are found
to satisfy the field equations under restriction on the
integration constants $g^{2}=\frac{1}{4\beta}$,
$V_{0}=\frac{g^{2}}{4A_{0}^{4}\phi_{0}^{2}}$and $C_{0}=0$. This
solution also leads to a power law inflation.

\section{Concluding remarks}
An excellent and remarkable feature of Noether symmetry has been
explored in the context of higher order gravity theory. Not only
the coupling parameters but also the solution of the field
equations can directly be obtained by applying Noether theorem in
such model. Earlier in a series of papers it has been shown that
for a unique and correct quantum description of higher order
gravity models, auxiliary variables should be chosen carefully in
a unique manner. The technique of choosing such auxiliary variable
now reveals new direction in the classical context also as Noether
symmetry has been found to be a powerful tool to explore solutions
to the field equations in highly nonlinear dynamics.

\end{document}